# A Novel Term_Class Relevance Measure for Text Categorization

D S Guru, Mahamad Suhil*

*Department of Studies in Computer Science, University of Mysore, Mysore, India*

**Abstract**

In this paper, we introduce a new measure called Term_Class relevance to compute the relevancy of a term in classifying a document into a particular class. The proposed measure estimates the degree of relevance of a given term, in placing an unlabeled document to be a member of a known class, as a product of Class_Term weight and Class_Term density; where the Class_Term weight is the ratio of the number of documents of the class containing the term to the total number of documents containing the term and the Class_Term density is the relative density of occurrence of the term in the class to the total occurrence of the term in the entire population. Unlike the other existing term weighting schemes such as TF-IDF and its variants, the proposed relevance measure takes into account the degree of relative participation of the term across all documents of the class to the entire population. To demonstrate the significance of the proposed measure experimentation has been conducted on the 20 Newsgroups dataset. Further, the superiority of the novel measure is brought out through a comparative analysis.

* Corresponding author.
 *E-mail address:* mahamad45@yahoo.co.in

## 1. Introduction

For the few decades automatic content based classification of documents from huge collections has become an active area of research due to the fact that electronic data over the internet has become unmanageably big and day by day it is increasing exponentially. Manual, tag based classification have lost their significance because of the huge size of the data that need to be processed and inability of the tags in describing the content of the documents. Varieties of applications of text classification which are of current demand such as spam filtering in E-mails, classification of E-Books, classification of news documents, classification of text data from social networks and so on have also made the researchers to explore various ways of analyzing and representing these data so that quick and efficient retrieval and management of this huge data can be done.

*1.1. A review of the available term weighting schemes*

As our work focuses on proposal of a new term weighting scheme, but not on classification framework, here we consider the literature on only different term weighting schemes.

Terms are the basic information units of any text document. So, all weighting schemes developed in the literature measure the weight of a term in representing the content of a document [1-5]. Based on whether the membership of the document in predefined categories is provided to measure the weight of a term or not, term weighting schemes are broadly classified into two classes namely, unsupervised term weighting schemes and supervised term weighting schemes. In the following subsections we provide a review of both the weighting scheme along with the techniques which have adopted them.

*1.1.1 Unsupervised term weighting schemes*

Most of the unsupervised term weighting schemes are from the information retrieval field. These methods are very useful when the training documents are not labeled by their class labels. The traditional term weighting methods borrowed from IR, such as binary, term frequency (TF), TF-IDF, and its various variants are unsupervised schemes [2]. The TF-IDF proposed by Jones [6, 7] and its variants are the most widely used term weighting schemes for text classification. Some of the variants of TF are Raw term frequency, log(TF), log(TF+1), or log(TF)+1[1-2]. If $n_i$ is the number of documents containing the term and N is the number of documents in the collection then, the variants of IDF are $1/n_i$, $\log(1/n_i)$, $\log(N/n_i)$, $\log(N/n_i)+1$ and $\log(N/n_i-1)$[1]. In [18], a novel inverse corpus frequency (ICF) based technique is proposed which computes the document representation in linear time.

*1.1.2 Supervised term weighting schemes*

Supervised term weighting schemes were developed especially for text categorization because of the fact that a supervised knowledge on the class labels of the training samples is provided [1-4]. All the supervised term weighting schemes make use of this class information in different ways. Supervised term weighting schemes are further classified into subcategories, based on whether the weight estimates relevancy of a term in preserving document content or the relevancy of a term in placing a document as a member of a class. So, it will be more effective to call the weighting schemes which are used to measure the relevance of a term in preserving the document content as term-document relevance measures and those which can be used to measure the term relevance in categorizing a document as term_class relevance measures.

**Term-Document Relevance measures:**

These measures are useful to select a discriminating subset of terms for representing a document by weighting the terms according to their relevance in preserving the content of the document. These are created by replacing the IDF component of the TF-IDF scheme. Most frequently used techniques to replace IDF include chi-square measure ($X^2$),

Information Gain(IG), Gain Ratio, Mutual Information(MI), Odds Ratio(OR) [1-4,8-12]. From past few years, many researchers have proposed alternative term-document relevance schemes [1, 13-16]. All these are basically feature selection techniques used in term weighting schemes. In [14], a comparison of corpus-based and class-based keyword selection is proposed by using TF-IDF as weighting scheme. In [4], a class-indexing-based term weighting for automatic text classification is proposed. An inverse class space density frequency ($ICS_\delta F$) is used along with TF-IDF method that provides a positive discrimination on infrequent and frequent terms.

**Term_class Relevance measures:**

These measures compute the ability of a term in classifying a document as a member of a class. To the best of our knowledge, only one work of this category has been proposed by Isa et al., [20] using Bayes posterior probability. Though, some works make use of Bayes probability for representation, they have not clearly stated the advantage of the measure in classification [11, 18]. After [20], this measure was extensively used for term weighting [21, 22]. The beauty of this measure lies in the fact that, instead of computing the weight of a term in preserving the content of a document, the relevancy of the term in categorizing the document as a member of a class can be measured directly. Which is computed as the Bayes posterior probability $P(C_j/ t_i)$ for a class $C_j$ and term $t_i$ as given by,

$$P(C_j / t_i) = \frac{P(t_i / C_j) P(C_j)}{P(t_i)}$$

where,

$$P(C_j) = \frac{Total\_of\_words\_in\_C_j}{Total\_of\_words\_in\_training\_dataset},$$

$$P(t_i) = \frac{\sum occurrence\_of\_t_i\_in\_all\_categories}{\sum occurrence\_of\_all\_terms\_in\_all\_categories}, \text{ and}$$

$$P(t_i / C_j) = \frac{occurence\_of\_t_i\_in\_C_j}{\sum occurrence\_of\_all\_terms\_in\_C_j}$$

To make use of the complete advantage of the proposed relevance measure, Isa et al., [20] also propose a text representation scheme which works with the reduced dimension for each document at the time of representation itself. This work happened to be the very first of its kind in the literature of text classification where, a document is represented only with number of dimensions equal to the number of classes in the corpus without any dimensionality reduction technique applied.

In this representation scheme, first, a matrix $F$ of size $m\,X\,k$ is created for every document where, m is the number of terms assumed to be available in the document and k is the number of classes. Then, every entry $F(i, j)$ of the matrix is filled by the relevancy of the corresponding term $t_i$ in classifying the corresponding document as a member of class $C_j$. Then, a feature vector f of dimension k is created as a representative for the document where, f(j) is the average of relevancy of every term to a class $C_j$. It shall be carefully observed here that, a document with any number of terms is represented with a feature vector of dimension equal to the number of classes in the population which is very small in contrast to the feature vector that is created in any other vector space representation scheme where the dimension is equal to the total number of terms due to all documents of the population. Therefore, a great amount of dimensionality reduction is achieved at the time of representation itself without the application of any dimensionality reduction technique. However, the classification accuracy accomplished is not of that high. Motivated by this work, in this paper we propose a novel term_class relevance measure with the following objectives,

- Exploiting the complete advantage of text representation scheme proposed by Isa et al.,[20].

- Comparison of the effectiveness of the proposed term_class relevance measure with that of Bayes posterior probability based measure.

- Isa et al., [20] make use of SVM as the classifier. So we are also investigating the effect SVM on our proposed relevance measure and also compare it with other available classifiers.

The rest of the paper is organized as follows. The proposed term_class relevance measure is presented in the section 2. In section 3, presents the results and discussion on the experimentation. A comparative analysis of the proposed relevance measure with other contemporary works is given in the Section 4. Finally, section 5 presents the conclusion and future enhancements.

## 2. A New Term_Class Relevance Measure

In this section, we propose a novel measure called term_class relevance measure. Term_class relevancy is defined as the ability of a term '$t_i$' in classifying a document 'D' as a member of a class '$C_j$'. We begin with introducing two new concepts which decide the role of a term in a class, namely, Class_Term Weight and Class_Term Density.

**Class_Term Weight:** It is the relative weight of the term with respect to a class of interest which is computed by counting only those documents of the class of interest that are containing the term of interest against that of the entire corpus. That is, the class_term weight of a term '$t_i$' in the class '$C_j$' is computed as the ratio of $ClassFrequency(t_i, C_j)$ to the $CorpusFrequency(t_i)$. It is given by the equation below.

$$Class\_TermWeight(t_i, C_j) = \frac{ClassFrequency(t_i, C_j)}{CorpusFrequency(t_i)}$$

where, $ClassFrequency(t_i, C_j)$ is the number of documents of '$C_j$' containing '$t_i$' at least once and $CorpusFrequency(t_i)$ is the number of documents of the entire corpus containing '$t_i$' at least once.

If the class_term weight of a term $t_i$ with respect to the class $C_j$ is very high then the probability that the document D which contains $t_i$ is most likely a member of the class $C_j$ is also high. Therefore, the relevancy of a term which we call it as $Term\_Class\,Relevancy(t_i, C_j)$ in deciding the class of a document is directly proportional to the class_term weight of the term. i.e.,

$$Term\_Class\,Relevancy(t_i, C_j) \propto Class\_TermWeight(t_i, C_j) \qquad (1)$$

**Class_Term Density:** It is the relative density of a term of interest with respect to the class of interest. It is computed as the ratio of the number of occurrences of the term in the class of interest to that of the entire corpus. That is, the class_term density of a term '$t_i$' with respect to the class '$C_j$' is computed as the ratio of frequency of $t_i$ in $C_j$ to its frequency in the corpus. It is given by the equation below.

$$Class\_TermDensity(t_i, C_j) = \frac{TermFrequency(t_i, C_j)}{\sum_{j=1}^{k} TermFrequency(t_i, C_j)}$$

where, $TermFrequency(t_i, C_j)$ is the frequency of '$t_i$' in the class '$C_j$' which is computed as the sum of the frequencies of '$t_i$' in every document of '$C_j$' as shown by the equation below.

$$TermFrequency(t_i, C_j) = \sum_{doc=1}^{d_j} Frequency(t_i, D_{doc})$$

where, $Frequency(t_i, D)$ is the frequency of occurrence of the term 't$_i$' in document D and d$_j$ is the number of documents in the class C$_j$.

It shall be noticed that, if the class_term density of a term 't$_i$' in a class C$_j$ is very high then the probability that a document D which contains 't$_i$' is most likely a member of the class C$_j$ is also high. Therefore, the relevancy of a term in deciding the class of a document is directly proportional to the class_term density of the term. i.e.,

$$Term\_Class\,Relevancy(t_i, C_j) \propto Class\_TermDensity(t_i, C_j) \qquad (2)$$

By combining (1) and (2), the term_class relevancy is directly proportional to the product of the class_term weight and class_term density of the term,

$$Term\_Class\,Relevancy(t_i, C_j) \propto Class\_TermWeight(t_i, C_j) * Class\_TermDensity(t_i, C_j)$$

i.e.,
$$Term\_Class\,Relevancy(t_i, C_j) = c * Class\_TermWeight(t_i, C_j) * Class\_TermDensity(t_i, C_j)$$

where, c is the proportionality constant, which we decide based on the class weight with respect to the entire population as described below.

**Class Weight** ( $c$ ): It is the weight of the j$^{th}$ class C$_j$ in the corpus which is computed as the ratio of the number of documents in C$_j$ denoted by $Size\_of(C_j)$ to the total number of documents in the entire corpus as given by,

$$ClassWeight(C_j) = \frac{Size\_of(C_j)}{\sum_{j=1}^{k} Size\_of(C_j)} \qquad \text{Where, k is the number of classes.}$$

If each class has equal number of documents, then the class-weight serves as a scaling factor in computing the relevance of a term and it increases or decreases the relevancy of a term to a class when the size of the class compared to the size of other classes is larger or smaller respectively.

Therefore, the proposed relevancy measure of a term t$_i$ in placing a document D as a member of a class C$_j$ is given by the product of the three aspects namely, Class weight, Class_Term weight and Class_Term Density as given by the formula below.

$$Term\_Class\,Relevancy(t_i, C_j) = c * Class\_TermWeight(t_i, C_j) * Class\_TermDensity(t_i, C_j)$$

The main advantages of the proposed term_class relevancy measure are as follows,

- It directly computes the relevancy of the term with respect to a class of interest; which can itself be used as a clue to identify the possible class to which a document may belong without the need of a classifier.

- The measure uses class as well as corpus information together as opposed to the conventional TF-IDF scheme, which utilizes the document frequency from only the corpus.

- It shall be observed that, the relevancy of a term to a class is high only if the three factors class_term weight, class_term density and class_weight are high. This helps in properly deciding the weight of a term without any bias towards a particular class, which in turn helps in deciding the class for a classifier.

Once the term_class relevance of all terms of the training set of documents is computed with respect to every class present, each training document is then represented using the representation scheme proposed by Isa et al., [20] as explained in section 1.1.2. A document is first represented as a matrix $F$ of size $m \: X \: k$, where, m is the number of terms assumed to be available in the document and k is the number of classes. Then, every entry $F(i,j)$ of the matrix is filled by the relevancy of the corresponding term $t_i$ with respect to the class $C_j$. Then, a feature vector f of dimension k is created as a representative for the document where, f(j) is the average relevancy of all terms with respect to a class $C_j$. The feature matrix of size $n \: X \: k$ thus created for the n training documents is used for learning process. A similar vector of k dimension is created for the test documents and given to the learning algorithm or a classifier for labeling. The process of training and testing the classifiers is explained in the next section.

## 3. Classification with SVM and k-NN classifiers

To evaluate the applicability of the proposed term_class relevance measure, we make use SVM as learning algorithm to perform classification because of its good generalization ability. Moreover, the training burden for SVM is very less even though, the time required for training is directly proportional to the training dataset, because the representative feature vectors are of dimension equal to the number of classes only. So, to test the effectiveness of the proposed relevance measure we have experimented with the SVM classifier with Linear, Gaussian radial basis function (RBF) and Polynomial kernels.

We consider the 20 Newsgroups data set for our experimentation. It consists of approximately 20,000 newsgroup documents consisting 20 classes with each class bearing nearly equal number of samples. It has become a popular data set for text classification and clustering applications. Some of the documents are closely related to each other while others are highly unrelated. We conduct experiments with various proportion of training set to validate the performance of the proposed relevancy measure.

Fig 1 shows the overall classification accuracy of the system with various percentages of training samples using SVM classifier with different kernels. Fig 2 shows the precision of the SVM classifier with different kernels and Recall is shown in Fig 3. In Fig 4, the overall F-measure is presented. It can be observed from the figures (1-4) that, the SVM classifier with RBF kernel is working well when compared to the other kernels. The results are also presented graphically in figures below.

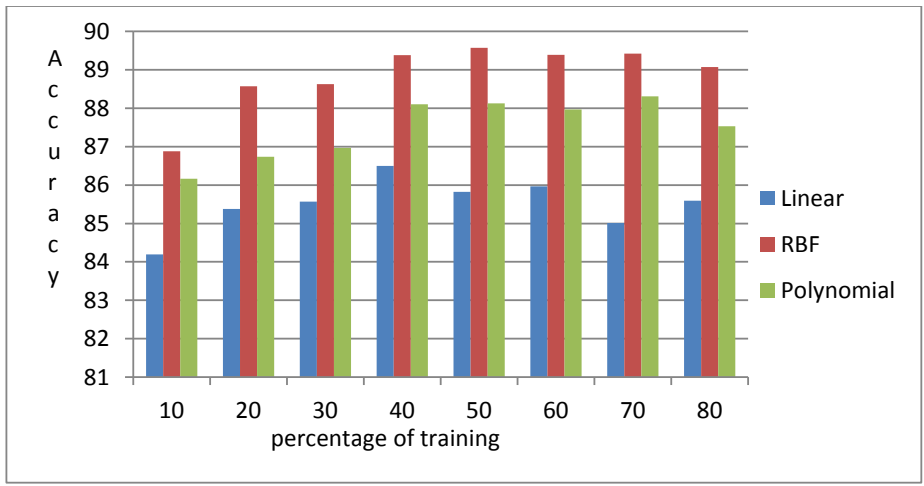

Fig 1. Overall Accuracy of classification with Linear, RBF and Polynomial kernels

*D.S. Guru and Mahamad Suhil / Procedia Computer Science 45 (2015) 13 – 22* 19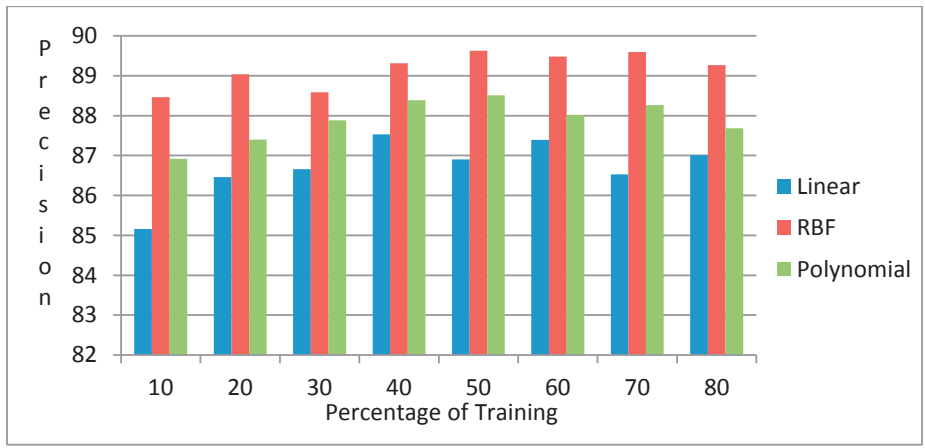

Fig 2. Precision of the SVM classifier with Linear, RBF and Polynomial kernels

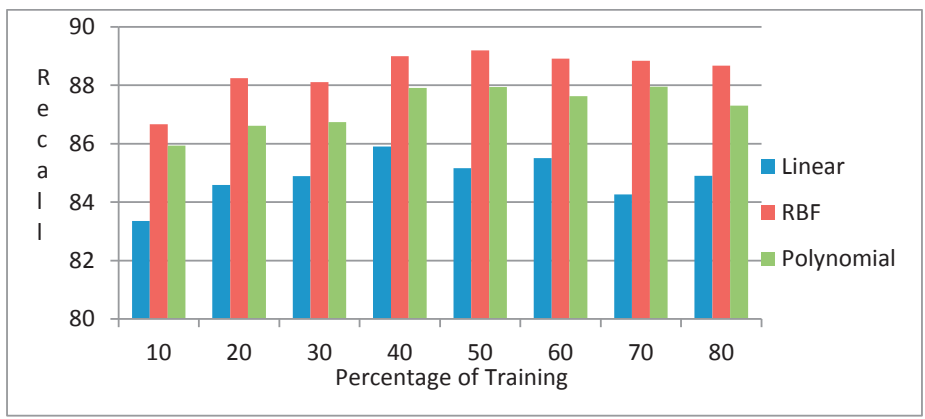

Fig 3. Recall of the SVM classifier with Linear, RBF and Polynomial kernels

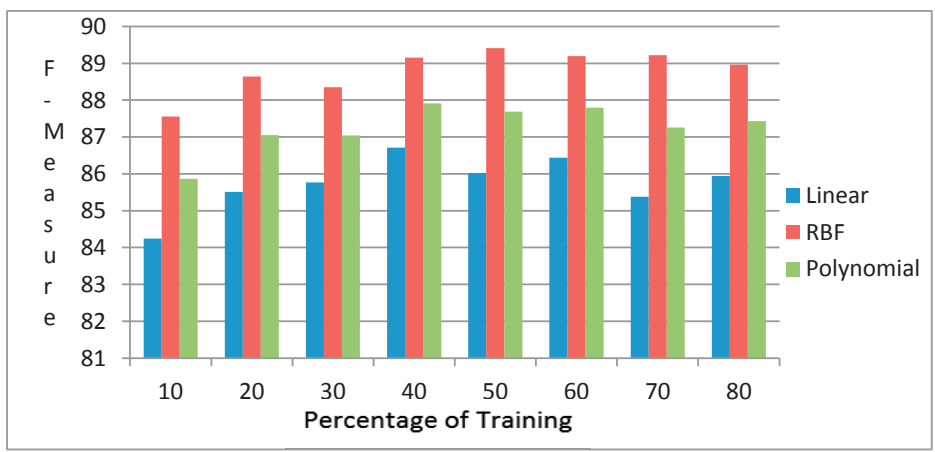

Fig 4. F-measure of the SVM classifier with Linear, RBF and Polynomial kernels

Further, the k-NN classifier is also adapted to test the proposed method because of its simplicity in classification. We performed the experimentation with various values of k from 1 to 20 and the performance of the classifier was high for k=10 with Euclidean distance being the proximity measure. Table 1 shows the results of k-NN classifier for k=10 and a comparison with the best results of SVM is also given. It can be observed that, k-NN has outperformed SVM. But, this performance is at the cost of computational complexity at the testing time.

Table 1. Results of SVM with RBF kernel and k-NN with k=10

| % of Trainig | Accuracy | | Precision | | Recall | | F-measure | |
|---|---|---|---|---|---|---|---|---|
| | k-NN | SVM | k-NN | SVM | k-NN | SVM | k-NN | SVM |
| 10 | **90.38** | 86.88 | **90.12** | 88.46 | **90.14** | 86.67 | **90.13** | 87.56 |
| 20 | **91.15** | 88.57 | **91.10** | 89.04 | **90.72** | 88.24 | **90.91** | 88.64 |
| 30 | **91.78** | 88.63 | **91.61** | 88.59 | **91.43** | 88.11 | **91.52** | 88.35 |
| 40 | **92.04** | 89.38 | **91.99** | 89.31 | **91.66** | 88.99 | **91.82** | 89.15 |
| 50 | **91.49** | 89.57 | **91.40** | 89.62 | **91.18** | 89.20 | **91.29** | 89.41 |
| 60 | **92.26** | 89.39 | **92.20** | 89.48 | **91.89** | 88.91 | **92.04** | 89.20 |
| 70 | **92.14** | 89.43 | **92.23** | 89.60 | **91.76** | 88.84 | **91.99** | 89.22 |
| 80 | **93.01** | 89.07 | **93.03** | 89.27 | **92.66** | 88.67 | **92.85** | 88.97 |

To compare the class-wise performance of each classifier we show the variation of F-measure vs. class in Fig 5 and 6. Fig 5, shows the values of F-measure vs. each class using k-NN classifier with k=10 and 10 percent of training. It can be noticed that, the performance is relatively low for classes 2, 3, 4, 7, 13 and 20. Further, the F-measure of SVM classifier vs. each class with RBF kernel and 10 % of training is shown in Fig 6. Though, the results of SVM are poor when compared to k-NN, SVM also has shown relatively low performance for the same classes as in the case of k-NN.

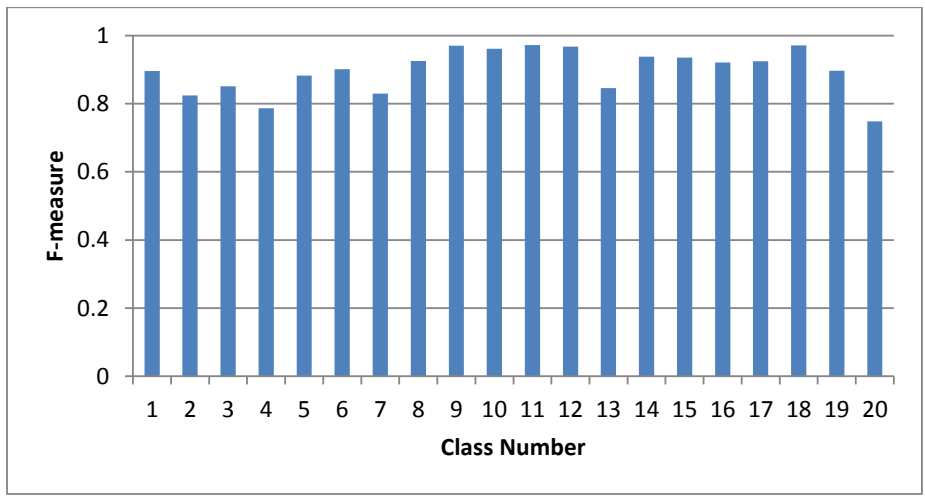

Fig 5. Classification performance vs. class for k-NN classifier with 10 % training

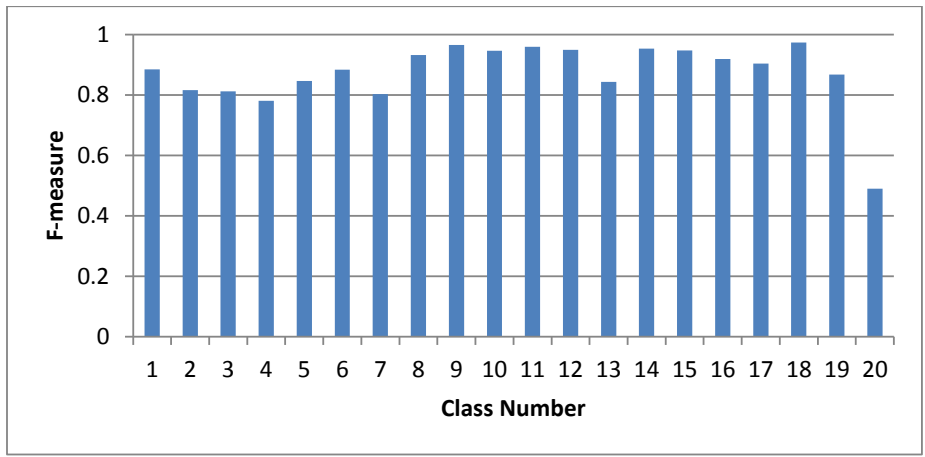

Fig 6. Classification performance vs. class for SVM classifier with RBF kernel and 10 % training

Table 2. Comparison of accuracy of the proposed with the work of Isa et al.,[20].

| Percentage of Training | Accuracy from [20] with SVM | | | Accuracy of Proposed Method | | | |
|---|---|---|---|---|---|---|---|
| | | | | SVM | | | k-NN |
| | Linear | RBF | Polynomial | Linear | RBF | Polynomial | |
| 30 | 83.04 | 82.97 | 82.83 | 85.57 | 88.63 | 86.97 | 91.78 |
| 70 | 88.02 | 87.88 | 87.93 | 85.02 | 89.43 | 88.31 | 92.14 |

## 4. Comparative Analysis

In this section, we provide a quantitative comparative analysis of the proposed term_class relevance measure with the results of Isa et al.,[20] in Table 2. The results corresponding to [20] have been extracted directly from the paper as the representation scheme is same in both the works and they also have provided the results on the same 20Newsgroups dataset using only SVM classifier with different kernels. Though there are a couple of works available in the literature, we are not comparing with any such works as they make use of the conventional vector space model for the representation of text documents. We can notice from the Table 2 that, the proposed term_class relevance measure outperforms the measure used by Isa et al.,[20]. Along with SVM, we compare the results using results of k-NN classifier with k=10. It can also be noticed from the Table 3 that, k-NN classifier with k=10 is showing enhanced results when compared to SVM with all the kernels for both the relevance measures. So, we recommend using k-NN as the classifier for better classification performance.

## 5. Conclusion

In this paper, a novel term_class relevance measure to compute the relevance of a term in classifying an unknown document as a member of a particular class is proposed. The proposed term_class relevance measure is a product of three aspects namely class_term weight, class_term density and class_weight. Experiments are conducted on 20 Newsgroups dataset using the SVM and k-NN classifiers. An effective text representation scheme which allows representation of text documents in reduced dimension is adapted to test the proposed term_class relevance measure. The comparative analysis of the results of the proposed work with the other contemporary research works shows the superiority of the proposed term_class relevance measure.

## 6. Acknowledgements

The second author of this paper acknowledges the financial support rendered by the University of Mysore UPE grants for the High Performance Computing laboratory.

## References

1. Lan, M., Tan, C. L., Su. J., and Lu, Y.2009. Supervised and Traditional Term Weighting Methods for Automatic Text Categorization. IEEE Transactions on Pattern Analysis and Machine Intelligence, Volume: 31 (4), pp. 721 – 735
2. G. Salton and C. Buckley. 1988. Term-Weighting Approaches in Automatic Text Retrieval, Information Processing and Management, vol. 24(5), pp. 513-523.
3. Debole F, Sebastiani. F. 2003. Supervised Term Weighting for Automated Text Categorization. Proceedings of the 2003 ACM symposium on applied computing, pp. 784-788.
4. Ren F, Sohrab M. G., 2013. Class-indexing-based term weighting for automatic text classification. Information Sciences 236 (2013) 109–125
5. Harish B. S., Guru D. S., and Manjunath. S. (2010). Representation and Classification of Text Documents: A Brief Review. IJCA Special Issue on "Recent Trends in Image Processing and Pattern Recognition" RTIPPR, pp. 110-119.
6. K. S. Jones, 1972. A statistical interpretation of term specificity and its application in retrieval, Journal of Documentation, Vol. 28, pp. 11-21.
7. K. S. Jones, 2004. A statistical interpretation of term specificity and its application in retrieval, Journal of Documentation, Vol. 60, pp. 493-502
8. Altınçay H, Erenel Z., 2010. Analytical evaluation of term weighting schemes for text categorization. Pattern Recognition Letters Vol. 31, pp-1310–1323.
9. Liu, Y., Loh, H.T., Sun, A., 2009. Imbalanced text classification: A term weighting approach. Expert Systems with Applications 36, 690–701
10. Mladenic, D., Grobelnik, M., 2003. Feature selection on hierarchy of web documents. Decision Support Syst. 35 (1), 45–87
11. Sebastiani, F., 2002. Machine learning in automated text categorization. ACM Comput. Surveys 34 (1), 1–47
12. Yang, Y., Pedersen, J.O., 1997. A comparative study on feature selection in text categorization. In: Proc. ICML'97, 14th Internat. Conf. on Machine Learning. Morgan Kaufmann Publishers, San Francisco, US, pp. 412–420
13. Liu, H., Yu, L., 2005. Toward integrating feature selection algorithms for classification and clustering. IEEE Trans. Knowledge Data Eng. 17 (4), 491–502
14. Ozgur, A., Ozgur, L., Gungor, T., 2005. Text categorization with class-based and corpus-based keyword selection. In: Proc. 20th Internat. Symp. on Computer and Information Sciences. Lecture Notes in Computer Science, vol. 3733, Springer-Verlag, pp. 606–615.
15. Tsai, R.T., Hung, H., Dai, H., Lin, Y., Hsu, W., 2008. Exploiting likely-positive and unlabeled data to improve the identification of protein–protein interaction articles. BMC Bioinform. 9
16. Wang, D, Zhang, H., 2013. Inverse-Category-Frequency Based Supervised Term Weighting Schemes for Text Categorization. Journal of Information Science and Engineering Vol 29, pp. 209-225
17. Reed, J, W., Jiao, Y., Potok T, E., Klump, B, A., Elmore, M, T., and Hurson, A, R., 2006. TF-ICF: A New Term Weighting Scheme for Clustering Dynamic Data Streams. 5th International Conference on Machine Learning and Applications. pp.258-263. IEEE Computer Society Washington
18. Fuhr, N., Hartmann, S., Lustig, G., Schwantner, M., Tzeras, K., Darmstadt, T. H., et al. (1991). AIR/X – A rule-based multistage indexing system for large subject fields. In: Proceedings of the proceedings of RIAO(pp. 606–623)
19. P. Soucy and G.W. Mineau, "Beyond tfidf Weighting for Text Categorization in the Vector Space Model," Proc. Int'l Joint Conf. Artificial Intelligence, pp. 1130-1135, 2005
20. Isa, D., Lee, L. H., Kallimani, V. P., and Raj Kumar, R. 2008. Text document preprocessing with the Bayes formula for classification using the support vector machine. IEEE Transactions on Knowledge and Data Engineering. Vol. 20, pp. 23 – 31.
21. Isa, D., Kallimani, V. P., Lee, L. H., 2009. Using the self-organizing map for clustering of text documents. Expert Systems with Applications. Vol. 36, pp. 9584–9591.
22. Guru D. S., Harish B. S., and Manjunath. S. 2010. Symbolic representation of text documents. In Proceedings of Third Annual ACM Bangalore Conference. doi. 10.1145/1754288.1754306.